\documentclass{article}
\usepackage[utf8]{inputenc}
\usepackage[margin=3cm]{geometry}
\usepackage{natbib}
\usepackage{amsmath, amssymb, graphicx, psfrag}
\usepackage{algorithm, algorithmic}
\usepackage{color}
\usepackage{url}

\usepackage{physics}
\usepackage{siunitx}
\usepackage{bm}

\usepackage{pgfplots} 
\usetikzlibrary{external}
\usepackage[bf,nooneline,tight,small,FIGTOPCAP]{subfigure}
\pgfplotsset{compat=newest} 
\pgfplotsset{plot coordinates/math parser=false,tick scale binop=\times,legend style={font=\footnotesize}} 
\newlength\figureheight 
\newlength\figurewidth

\title{Population Calibration using Likelihood-Free Bayesian Inference}
\author{Christopher Drovandi$^*$$^\dagger$$^\ddagger$$^\diamond$, Brodie Lawson$^\dagger$$^\ddagger$$^\diamond$$^\triangleright$, \\ Adrianne L Jenner$^\dagger$$^\ddagger$$^\diamond$ and Alexander P Browning$^\dagger$$^\ddagger$$^\diamond$ \\ \\ $^\dagger$School of Mathematical Sciences, Queensland University of Technology  \\ $^\ddagger$Centre for Data Science, Queensland University of Technology \\ $^\diamond$ARC Centre of Excellence for Mathematical and Statistical Frontiers \\
$^\triangleright$ARC Centre of Excellence for Plant Success in Nature and Agriculture \\	
	 \\ *corresponding author: c.drovandi@qut.edu.au \\ \\
}
\date{\today}

\begin{document}
	
	\setlength{\parindent}{0pc}
	\setlength{\parskip}{1ex}

	\maketitle
	
	\begin{abstract} 
	
	In this paper we develop a likelihood-free approach for population calibration, which involves finding distributions of model parameters when fed through the model produces a set of outputs that matches available population data.  Unlike most other approaches to population calibration, our method produces uncertainty quantification on the estimated distribution.  Furthermore, the method can be applied to any population calibration problem, regardless of whether the model of interest is deterministic or stochastic, or whether the population data is observed with or without measurement error.  We demonstrate the method on several examples, including one with real data.  We also discuss the computational limitations of the approach.  Immediate applications for the methodology developed here exist in many areas of medical research including cancer, COVID-19, drug development and cardiology.
	
	\end{abstract}

	\section{Introduction}
	\label{sec:intro}
	
	In this paper we are interested in the problem of determining the distribution of inputs for  a black-box function (deterministic or stochastic) that, when passed through this function, produces a distribution of outputs that `matches' some target distribution. The most immediate use-case for this process is in calibrating a mathematical model to data that exhibits heterogeneity, where this heterogeneity is important. For example, it is insufficient to consider only the ``mean'' individual when analysing the absorption dynamics of a drug treatment~\citep{Rieger2018} or the potential for life-threatening side-effects~\citep{Passini2017}. Instead, we wish to learn the spread of hidden model parameters across the whole population, as implied by the associated spread of measurable properties. Performing this calibration in a statistically rigorous way can be critical to learning population parameters that are actually representative of the true dynamics~\citep{Lawson2018}.
	
	Biological variability is exhibited at all levels of living organisms and readily captured by data \citep{Britton2013}. This data can be measurements which compare cell variability or differences in disease dynamics across the human population. For example, an individual's cancer growth and treatment is extremely heterogeneous and this heterogeneity leads to the minimal success of most drugs at clinical trials \citep{burrell2014tumour,fisher2013cancer}. The variability in human disease responses has also been evident in the COVID-19 pandemic, where hospitalised patient biomarkers illustrate the extensive variability in human immune responses to SARS-CoV-2 infection \citep{mathew2020deep,lucas2020longitudinal}. With all this available data, it is crucial for a method to be developed that can learn model population parameters capturing the heterogeneity in the data so as to better inform decision making.
	
	A variety of names have been given to this class of problem. Working in signal processing, \citet{Baggenstoss2017} calls it {\it pdf projection}. Practical approaches taken by more application-focused works refer to constructing {\it virtual populations}~\citep{Rieger2018,brown2015trauma,fuertinger2018virtual,jenner2021covid,jenner2021silico,barish2017evaluating} or {\it populations of models}~\citep{Britton2013,Passini2017}, owing to the intent to then carry out further computer experiments on those populations. Works in cell biology speak directly of identifying or estimating population heterogeneity~\citep{Hasenauer2011,Lambert2021}. Here, we suggest that the term \emph{population calibration} is suitable, highlighting that this is a calibration (inverse) problem, but one that respects the variability evident in a population. In Section~\ref{sec:background}, we formalise the problem of population calibration, and provide a short review of previous methodological approaches.
	
	In general, the population calibration problem is ill-posed; that is, there are potentially an infinite number of input distributions that can match the output distribution, or perhaps none at all in the case of model misspecification. Predominantly, previous approaches to population calibration return only a single distribution, and do not attempt to quantify the uncertainty in their result. We suggest that uncertainty quantification is important in this context, as it can provide insight into the parameters of the model whose distribution is well constrained by the population data. The most related method to ours is \citet{Hasenauer2011}, who also produce uncertainty quantification on the estimated input distribution using a similar approach. However, their method is developed for deterministic models with a known noise component.
	
	The main contribution of this paper is to develop a general approach for population calibration (Section \ref{sec:method}) that can, in principle, be applied in any context, including deterministic or stochastic models, and with or without extrinsic noise.  Further, in the presence of a stochastic model, it need not have a tractable likelihood function.  Crucially, our method quantifies the uncertainty in the estimated input distribution due to ill-posedness of the problem, and also a finite sample size of population data.  Our approach exploits the framework of likelihood-free inference~\citep{sisson2018}, which was originally designed to perform, or approximate, standard Bayesian inference for a statistical model with an intractable likelihood function. We have previously applied our likelihood-free approach to a specific population calibration problem in flow cytometry~\citep{Browning2021}, but here we formalise and generalise the method. By presenting our general framework, we also seek to bring to the attention of statisticians the population calibration problem and its sub-problems, some of which have been considered in the statistical literature but many others not.  A further contribution is that we discuss in detail the limitations of the general approach throughout the results and discussion sections. In particular, our quest for uncertainty quantification significantly increases the computational complexity of the problem.  We hope that this will inspire the development of new and more computationally efficient methods.

	\section{Background} \label{sec:background}
	
	The marginal distributions of two correlated random variables $\mathsf{X} \in \mathcal{X}$ and $\mathsf{Y} \in \mathcal{Y}$ are related by the law of total probability,
	\begin{equation}
	\label{base_formulation}
	h(y) = \int_{\mathcal{X}} g(y|x)f(x)\,dx
	\end{equation}
	where $f(x)$ and $h(y)$ are their marginal distributions, and $g(y|x)$ is the conditional density of $\mathsf{Y}$ given $\mathsf{X}$. In the context of this paper, $g(y|x)$ defines the model with parameter $x$; for a deterministic function $y = \mathcal{F}(x)$, the conditional density is $g(y|x) = \delta(y - \mathcal{F}(x))$ where $\delta$ is the Dirac delta function, whereas for a stochastic model, $g(y|x)$ is the conditional density of the model output given the parameter. The fundamental population calibration problem is to find $f(x)$, given $h(y)$ and $g(y|x)$.
	
	Equation~\eqref{base_formulation} is the Fredholm equation of the first kind, and where $g(y|x)$ can be evaluated pointwise for a given $x$ and $y$, density deconvolution approaches may be used (see \cite{Crucinio2021} and references therein). These approaches are common in image processing~\citep{Lu2010} and have also been used, for example, to ``undo'' the stochastic delay between disease incidence and the observable event, death~\citep{Goldstein2009}. The expectation maximisation approaches often taken here do not naturally address the ill-posedness of the problem, however, requiring convolution with smoothing kernels to avoid concentration of $f(x)$ onto spikes~\citep{Crucinio2021}. 
	
	Methods for population calibration of deterministic models have also proliferated separately, without acknowledgement of the connection to Fredholm equations. Some approaches, particularly in cardiac electrophysiology, simplify the problem by matching to ranges of the population data rather than their distribution~(for example \cite{Britton2013, Passini2017}). \cite{brown2015trauma} infer parameters characterising the response to trauma in a small experimentally-observed population responding, then create a virtual population of 10,000 by sampling uniformly over the resultant parameter ranges.
	
	Efforts to statistically improve population calibration in these applied contexts have included capturing the distributional moments of the data~\citep{Tixier2017} or sampling from the data density then using a post-processing step to account for the missing volume factor that we discuss further below~\citep{Lawson2018}. \citet{allen2016efficient} generate a virtual population using data for cholesterol measurements. Virtual individual parameters are sampled from plausible ranges and then the virtual patient was included in the virtual population with some probability based on the original data distribution. 
	
	\citet{Hasenauer2011} take a unique approach and represent the unknown distribution function $f(x)$ as a finite mixture of overlapping kernel distributions (i.e., Gaussian), transforming the population calibration problem to an approximate high-dimensional inference problem. The primary advantage of this approach is that the samples from the output distribution $h(y)$ can be drawn using importance sampling with the mixture weights. While this approach allows the uncertainty in inferences to be quantified through uncertainty in individual mixture weights, the approach suffers from the requirement to pre-specify the locations and variances of the kernel functions, which may be difficult to do for a moderate to large number of model parameters.
	
	Several approaches have used the idea of entropy maximisation to regularise the population calibration problem's ill-posedness~\citep{Tixier2017, Dixit2020}. Maximising the entropy of $f(x)$ avoids the artificial introduction of information beyond what the data (or prior) provides, and hence avoids the over-concentration of $f(x)$ onto individual promising regions where others might also exist.
	
	By observing that equation~\eqref{base_formulation} can be written in reverse, and introducing a prior distribution $\pi_0(x)$ to make the formulation more general, we have that
	\[
	\begin{aligned}
	f(x) &= \int_{\mathcal{Y}} g^{-1}(x|y)h(y)\,dy \\
	&= \int_{\mathcal{Y}} \frac{\pi_0(x) g(y|x) }{\int_{\mathcal{X}} \pi_0(x') g(y|x') \, dx'}h(y)\,dy.
	\end{aligned}
	\]
	For a deterministic model $g(y|x) = \delta( \mathcal{F}(x) - y)$, this then simplifies to
	\begin{equation}
	\label{volume_corrected}
	f(x) = \frac{\pi_0(x)}{\int_{\mathcal{X} \in \mathcal{F}^{-1}(y)} \pi_0(x') \, dx'} h\bigl( \mathcal{F}(x) \bigr),
	\end{equation}
	as separately observed by~\citet{Baggenstoss2017} and~\cite{Lambert2021}. The $\left[\int_{\mathcal{X} \in \mathcal{F}^{-1}(y)} \pi_0(x') \, dx'\right]^{-1}$ term is a volume correction term akin to the Jacobian determinants seen for invertible transformations of random variables. This integration over the level sets of ${\cal F}$ inspired the terminology in those works of ``uniform manifold sampling'', and ``contour Monte Carlo'', respectively. \citet{Baggenstoss2017} demonstrates that entropy maximisation can be achieved by choice of the prior according to an energy statistic (and notably, that a uniform prior naturally provides maximum entropy solutions for bounded $\mathcal{X}$). \citet{Lambert2021} observe that the volume correction term can be approximated offline by repeatedly sampling from the $y$-marginal of $g(y|x)\pi_0(x)$, and estimating the density based on the resultant $y$ values using non- or semi-parametric density estimation
	\[
	q(y) \propto  \int_{\mathcal{X} \in \mathcal{F}^{-1}(y)} \pi_0(x') \, dx'.
	\] 
	So long as this density is sufficiently well-approximated,~\eqref{volume_corrected} can then be sampled from using any posterior sampling approach that requires only that the posterior be defined up to a normalisation factor.
	
	Typically, one does not have $h(y)$ in closed form, but instead a series of samples $\mathsf{y} = (y_1, y_2, \ldots, y_n)$ from which the density $h(y)$ must be estimated. This does also offer a hierarchical formulation, assigning individual parameter values $x_i$ to each observation and then regularising the distribution of $x$ values across the population via appropriate choice of hyperprior (for example~\cite{Alahakoon2021}). This type of approach may regain a traditional likelihood for population calibration, but could drastically increase the dimensionality of the sampling problem for the general population calibration problem. This therefore does not seem to be a scalable approach, at least when the data can be reasonably treated as independently and identically distributed samples from some distribution $h(y)$.  Furthermore, most approaches treat the estimated $h(y)$ as being known, and hence do not account for the uncertainty due to sample size, which could be considerable if $n$ is small.
	
	Several gaps remain in the literature surrounding population calibration. Although regularisation via entropy can help improve the single distribution $f(x)$ obtained by these different approaches, almost all return only a single distribution, with no concept of the sensitivity or certainty of this result. This is a significant downside, as uncertainty arises both due to the ill-posedness of the problem and the reliance upon estimation of $h(y)$ from a finite amount of data. Furthermore, most of the discussed approaches rely upon the model explicitly defining $g(y|x)$. In the approach we put forward here, one needs only to be able to sample from $g(y|x)$, and learns a {\it distribution} over distributions, unlocking the full benefits of the Bayesian paradigm for population calibration.

	\section{Likelihood-Free Bayesian Inference for Population Calibration} \label{sec:method}

	Here we describe our general population calibration approach, using the framework of likelihood-free inference.  We first provide an overview of likelihood-free inference in the context of a more standard Bayesian calibration problem.  We then show how the likelihood-free framework can be leveraged to solve the general population calibration problem, and importantly how it can produce uncertainty quantification on the estimated $f(x)$.

	\subsection{Likelihood-Free Bayesian Inference}
	
	For ease of exposition, we focus here on the popular methods of approximate Bayesian computation (ABC) and Bayesian synthetic likelihood (BSL). We note however that virtually any likelihood-free inference method could be considered for population calibration, and we refer to \citet{sisson2018} and \citet{Cranmer2019} for a more comprehensive overview of statistical and machine learning approaches, respectively.
	
	The goal of Bayesian inference is to estimate the posterior distribution of model parameters $x$, in light of the prior distribution $\pi(x)$ and the likelihood function for observed data, $g(\mathsf{y}|x)$:
	\begin{align*}
	\pi(x|\mathsf{y}) \propto g(\mathsf{y}|x)\pi(x).
	\end{align*}
	Likelihood-free inference methods have been developed for so-called implicit models, which are generative models that can be simulated but the corresponding likelihood function is computationally intractable and standard Bayesian inference approaches are thus ruled out.  ABC proposes to target the following approximate posterior
	\begin{align}
	\pi_\epsilon(x|\mathsf{y}) &\propto \pi(x) \int_{\mathcal{Y}} \mathbb{I}(\rho(\mathsf{y}, \mathsf{z}) < \epsilon) g(\mathsf{z}|x) d \mathsf{z}, \label{eq:abc_target}
	\end{align}  
	where $\mathsf{z} \in \mathcal{Y}$ is a dataset simulated from the model at $x$.  Here, $\rho(\mathsf{y}, \mathsf{z})$ is a function measuring the distance between observed and simulated data, and $\epsilon$ is a threshold defining what is `close enough', setting the indicator function $\mathbb{I}(\cdot)$ to one when the simulated and observed data are sufficiently similar and zero otherwise. It is rather common, but not essential, to compute the distance between observed and simulated datasets on the basis of a carefully chosen summary statistic function $S(\cdot):\mathcal{Y} \rightarrow \mathcal{S}$ so that $\rho(\mathsf{y}, \mathsf{z})$ becomes $\rho(S(\mathsf{y}), S(\mathsf{z}))$.  The primary motivation for summarising the data is to avoid measuring closeness between large datasets, which may be difficult to do efficiently.  However, comparing the empirical distribution of observed and simulated datasets can be effective, see \citet{Drovandi2021} for a review of such approaches.
	
	The integral in \eqref{eq:abc_target} can be estimated by Monte Carlo integration, drawing independent simulations of $ \mathsf{z}$ from $g$ for a given $x$.  Thus sampling of $\pi_\epsilon(x|\mathsf{y})$ can proceed without ever evaluating $g(\mathsf{y}|x)$.  There exist a plethora of algorithms for approximate sampling of $\pi_\epsilon(x|\mathsf{y})$, with common sampling classes including, but not limited to, rejection (e.g.\ \citet{Beaumont2002}), Markov chain Monte Carlo (MCMC, e.g.\ \citet{Marjoram2003}) and sequential Monte Carlo (SMC, e.g.\ \citet{Sisson2007}) sampling.  The accuracy of ABC hinges on the informativeness of the observed data summarisation $S(\mathsf{y})$ and the size of the tolerance $\epsilon$, noting that further computational burden is often required to improve the approximation with respect to either of these two aspects (i.e.\ by increasing the dimension of $S(\mathsf{y})$ or reducing $\epsilon$).     
	
	BSL is an alternative statistical approach for likelihood-free inference.  Whereas ABC can be interpreted as using a non-parametric approximation of the summary statistic likelihood $g(S(\mathsf{y})|x)$ \citep{blum2010approximate}, BSL uses a parametric approximation.  Specifically, \citet{wood2010} proposes that
	\[ g(S(\mathsf{y})|x) \approx \mathcal{N}(S(\mathsf{y}); \mu(x), \Sigma(x)),
	\]
	called the synthetic likelihood, where the mean $\mu$ and covariance $\Sigma$ may depend on the model parameter $x$.  Since such relationships are generally unknown, they are estimated empirically using sample moments calculated from the summary statistics of $m$ independent simulations of the model with parameter value $x$.  \citet{price2018bayesian} examine the synthetic likelihood in the Bayesian framework.  The estimated synthetic likelihood is often placed within an MCMC algorithm to sample the approximate posterior.
	
	BSL has the advantage over ABC in terms of the number of tuning parameters; it requires specifying $m$ the number of simulated datasets for estimating synthetic likelihood, for which there is guidance in \citet{price2018bayesian}.  Furthermore, it has been shown empirically \citep{price2018bayesian} and theoretically \citep{frazier2019bayesian} that BSL scales more efficiently than ABC when the dimension of the summary statistic increases, at the expense of the Gaussian assumption.  There are several extensions to BSL, such as a semi-parametric approach to relax the Gaussian assumption \citep{an2018robust}, whitening transformations of the summary statistic to reduce $m$ \citep{priddle2019efficient}  and handling model misspecification \citep{frazier2019robust}. However, for simplicity, we only consider standard BSL in this paper.

	\subsection{Our Approach}
	
	We now provide details on how we can leverage the likelihood-free framework to solve population calibration problems, and simultaneously quantify the uncertainty in the solution. We first assume that the true $f(x)$ belongs to a family of parametric distributions $f_\theta(x)$, parameterised by $\theta$.  We are interested in inferring $\theta$, particularly via its posterior distribution $\pi(\theta|\mathsf{y})$ in light of the population data $\mathsf{y}$ and a prior distribution $\pi(\theta)$ so that we can assess the uncertainty in the estimated $f(x)$.  For example, if $f_\theta(x)$ is the density function of a Gaussian distribution, then $\theta$ consists of a mean vector and covariance matrix, and the choice of prior might be something like a Gaussian-inverse-Wishart distribution.  It may seem restrictive to only search within a given family of distributions, but we point out that such a choice also naturally introduces controllable regularisation to address the generally ill-posed nature of the population calibration problem.  Posterior predictive checks can always be used to assess if the population data is well recovered, and if it is not, a more expressive family for $f_\theta(x)$ can be considered.   Domain expert knowledge could also be used to inform the choice of family for $f_\theta(x)$.
	
	We note that there may be other parameters to estimate that are assumed to not vary between individuals.  These may be parameters within the mathematical model (i.e.\ some components of $x$) that are assumed fixed for each individual or nuisance parameters such as the standard deviation of observation error.  We denote these parameters as $\phi$.

	\begin{figure}[!t]
		\centering
		
		\includegraphics[scale=0.75]{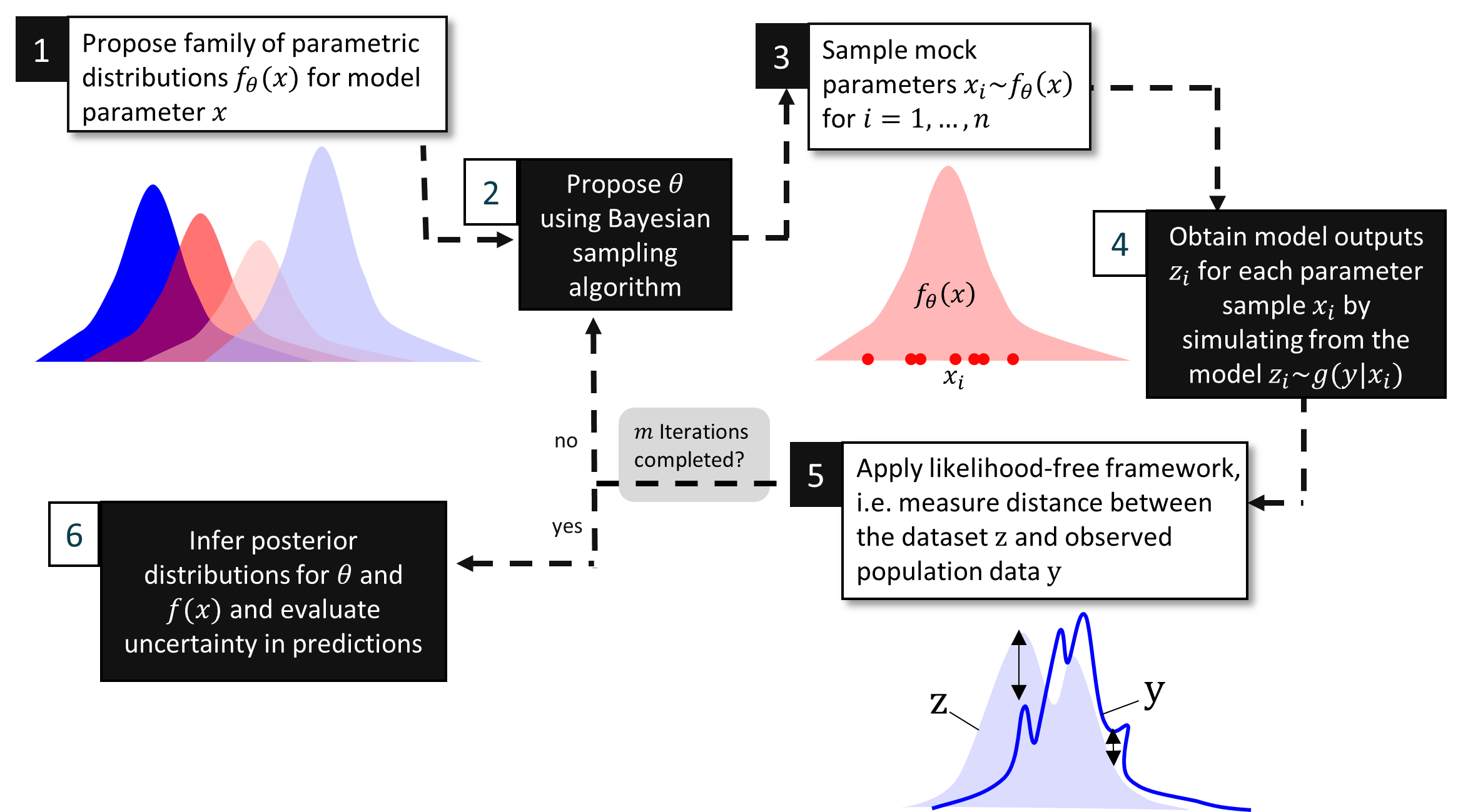}
		\caption{Consider heterogeneous parameters in the model $x$ are described by their unknown distribution $f(x)$ which we wish to estimate. We propose that the true $f(x)$ belongs to a family of parametric distributions $f_\theta (x)$, parameterised by $\theta$. For a given value of $\theta$, proposed from a Bayesian sampling algorithm, we sample mock parameters $x_i\sim f_\theta (x)$ and then simulate a mock population dataset $\mathsf{z}$ which is obtained by simulating the model for each mock parameter $x_i$, i.e., $z_i \sim  g(y|x_i,\phi)$ (or $z_i=g(x_i,\phi)$ in the deterministic case). We then employ the likelihood-free framework to measure the distance between the mock population dataset $\mathsf{z}$ and the observed data $\mathsf{y}$. This process of sampling $\theta$, simulating the mock population data set and comparing to the observed data is iterated $m$ times. After which we can infer posterior distributions for $\theta$ and $f(x)$ and evaluate uncertainty in our predictions. }
		\label{fig:approach_schematic}
		
	\end{figure}

	For a given value of $(\theta,\phi)$ proposed in a Bayesian sampling algorithm, we can simulate a mock population dataset $\mathsf{z}$ by drawing $x_i \sim f_\theta(x)$ and then simulating from the model, $z_i \sim g(y|x_i,\phi)$ (or $z_i = g(x_i,\phi)$ in the deterministic case) for $i = 1,\ldots,n$ such that $\mathsf{z} = (z_1,\ldots,z_n)$.  Then, all that is required to employ the likelihood-free framework is to devise an appropriate way to measure the distance between the datasets $\mathsf{z}$ with $\mathsf{y}$ in a distributional sense. That is to say, observations in each dataset are treated as unpaired and without ordering. For a summary of our approach see Figure  \ref{fig:approach_schematic}.

	There are several options for the comparison between $\mathsf{z}$ and $\mathsf{y}$. If the population data is low-dimensional, then it would be feasible to construct a distance function for ABC by directly comparing the empirical distributions of $\mathsf{x}$ and $\mathsf{y}$.  Some such distances reviewed in \citet{Drovandi2021} include the Wasserstein, Cramer von Mises and energy distances.  For higher dimensional population data, it may be more efficient to work with features of the distribution, such as its location, scale and dependence.  These summary statistics could be used in ABC, by choosing some appropriate distance function between summaries, or BSL. Regardless of the specific likelihood-free approach adopted, we compare simulated and observed datasets of the same size and we explicitly propagate the uncertainty due to final sample size through the estimation of $f(x)$. There is also flexibility with our approach. If there is no interest in accounting for uncertainty due to sample size it is possible to simulate datasets larger than the observed datasets. Conversely, if model simulation is expensive and the population data is large, then we can simulate less data than what was observed.  We now illustrate our approach on a variety of population calibration problems.

	\section{Examples} \label{sec:Examples}
	
	Code to produce the results are available at\\ \url{https://github.com/cdrovandi/Likelihood-Free-Population-Calibration}.
	
	\subsection{Mixture model}
	
	Here we consider the Fredholm integration problem of the first kind from \citet{Ma2011} (also considered in \citet{Crucinio2021}) with the following specifications:
	\begin{align*}
	f(x) &= \frac{1}{3}\mathcal{N}(x;0.3,0.015^2) + \frac{2}{3}\mathcal{N}(x;0.5,0.043^2), \\
	g(y|x) &= \mathcal{N}(y;x,0.045^2),
	\end{align*}
	which yields the output distribution
	\begin{align*}
	h(y) &= \frac{1}{3}\mathcal{N}(y;0.3,0.045^2 + 0.015^2) + \frac{2}{3}\mathcal{N}(x;0.5,0.045^2 + 0.043^2).
	\end{align*}
	This is a de-noising problem, the goal being to recover the input distribution $f(x)$ from the noise-corrupted output. Treating the noise corruption process as the model, however, it can also be interpreted as a population calibration problem for which both $x$ and $y$ have known, analytical expressions. As we go on to demonstrate, the significant amount of noise in this problem also acts to highlight its ill-posedness, and the importance of uncertainty quantification in this context.
	
	\begin{figure}[t]
		\centering
		
		\subfigure[]{\includegraphics{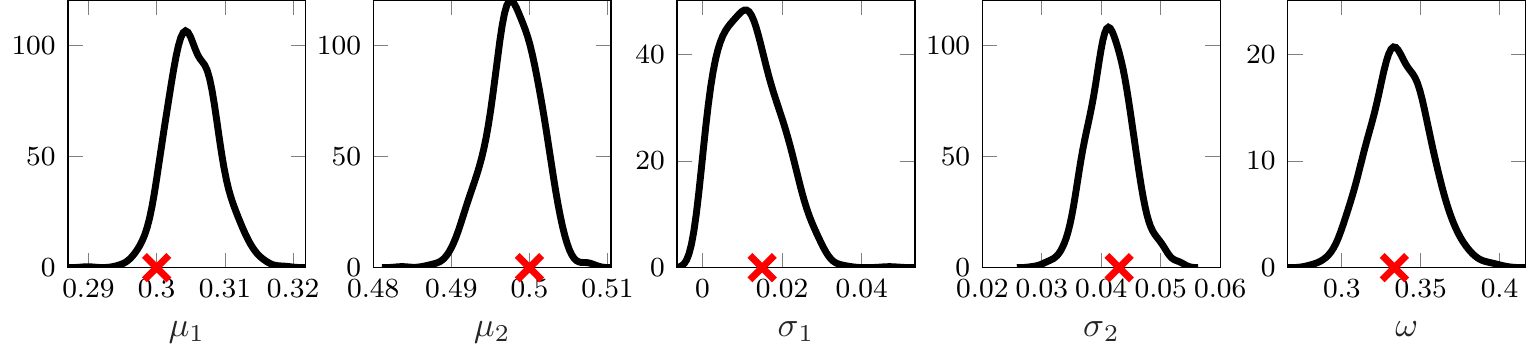}\label{figsub:mixture_posterior_theta}}
		\subfigure[]{\includegraphics{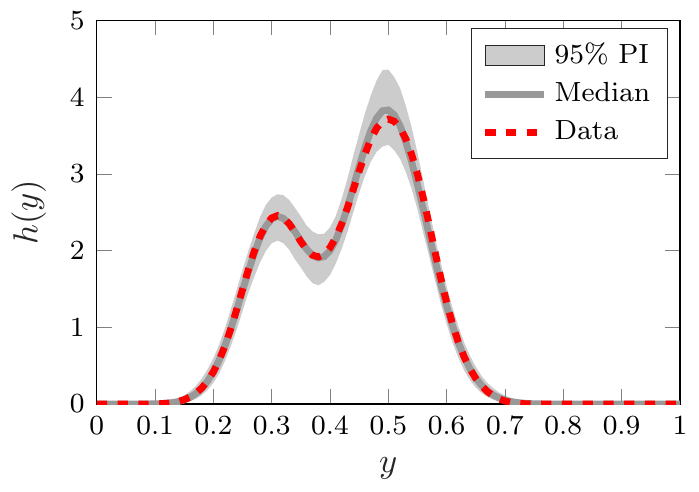}\label{figsub:mixture_posterior_output}}
		\subfigure[]{\includegraphics{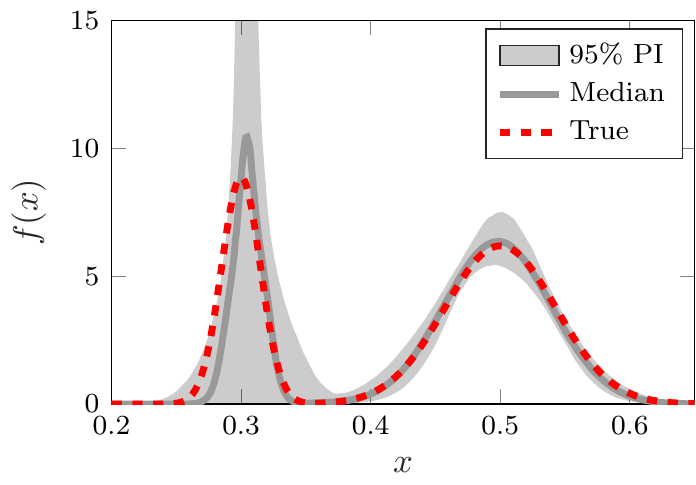}\label{figsub:mixture_posterior_input}}

		\caption{Results for the mixture model example.  (a) Univariate posterior distributions of $\theta$ with true values shown. (b) Posterior predictive distribution (median and 95\% PI) of the $h(y)$ with the true $h(y)$ overlaid (note here that kernel density estimates are used from the $n=1000$ samples of $h(y)$).  (c) Posterior distribution (median and 95\% PI) of $f(x)$ with the true $f(x)$ overlaid.}
		\label{fig:mixture_example}
		
	\end{figure}
	
	We assume that $n=1000$ observations $\mathsf{y} = (y_1,y_2,\ldots,y_n)$ have been drawn from $h(y)$ to attempt to infer $f(x)$.  As summary statistics we use the score of a two-component mixture of Gaussians, creating a five-dimensional summary statistic.  We assume that the parametric form of $f(x)$ is correctly specified. That is, it is specified as a two-component Gaussian mixture with $\theta = (\mu_1,\mu_2,\sigma_1^2,\sigma_2^2,\omega)$ to infer, where $\mu_i$ and $\sigma_i^2$ are the mean and variance of the $i$th component, and $\omega$ is the mixing weight for the first component.  Our approach approximates the posterior of $\theta$.  For priors we use $\mu_i \sim \mathcal{N}(0,1)$, $\sigma_i^2 \sim \mathcal{E}xp(1)$ and $\omega \sim \mathcal{U}(0,1)$.  In the prior we also impose the identifiability constraint that $\mu_1 < \mu_2$.  We use $10^5$ iterations of BSL with $m=50$ to sample the approximate posterior of $\theta$.  
	
	The results are shown in Figure \ref{fig:mixture_example}.  Figure \ref{fig:mixture_example}(a)   shows the univariate posterior distributions of $\theta$ with true values included.  It is evident that $\theta$ is well-identified, except for $\sigma_1$ where very small values are retained in the posterior.  Figure \ref{fig:mixture_example}(b) shows the posterior predictive distribution (specifically the median and 95\% interval) of the estimated $h(y)$.  Each prediction of $\mathsf{y}$ is summarised by a kernel density estimate (KDE), which is also applied to the observed data. The median of the predicted KDEs is similar to the KDE of the observed data\footnote{Strictly this point estimate is not a valid density estimate itself. Such a point estimate could be generated from a point estimate of $\theta$ or from finding the closest posterior KDE sample to the median point estimate.}.  Further, the 95\% interval of the predicted KDEs tightly enclose the observed KDE.  BSL is thus successful in finding combinations of $\theta$ that parametrically describe populations that exhibit the correct distribution of model output.

	Figure \ref{fig:mixture_example}(c) shows features of the posterior distribution of $f(x)$.  A fair amount of uncertainty in $f(x)$ is immediately seen, despite the close match to the output data across the posterior (Figure~\ref{fig:mixture_example}(b)). There is therefore a wide range of distributions $f(x)$ that all represent populations that could have produced the observed data --- the problem is indeed ill-posed. The uncertainty in $f(x)$ comes mainly from the uncertainty in $\sigma_1$, for which very small values are permitted. This occurs because in the observed data, variability from the second mixture component and from the noisy model $g(y|x)$ easily masks the effects of the first component's small variance. Uncovering these aspects of a population calibration problem is only possible through approaches such as ours, that go beyond returning a single distribution $f(x)$ as an answer. Of course, where a single answer is desirable, this is easily achieved once a posterior on distributions $f(x)$ has been obtained --- here, the median distribution is seen to agree well with the true population distribution. One could also choose the maximum entropy distribution from the posterior, so as to minimise any introduction of artificial knowledge~\citep{Tixier2017, Dixit2020}.

	\subsection{Deterministic growth factor model}

	\begin{figure}[!htp]
		\centering
		
		\subfigure[]{\includegraphics{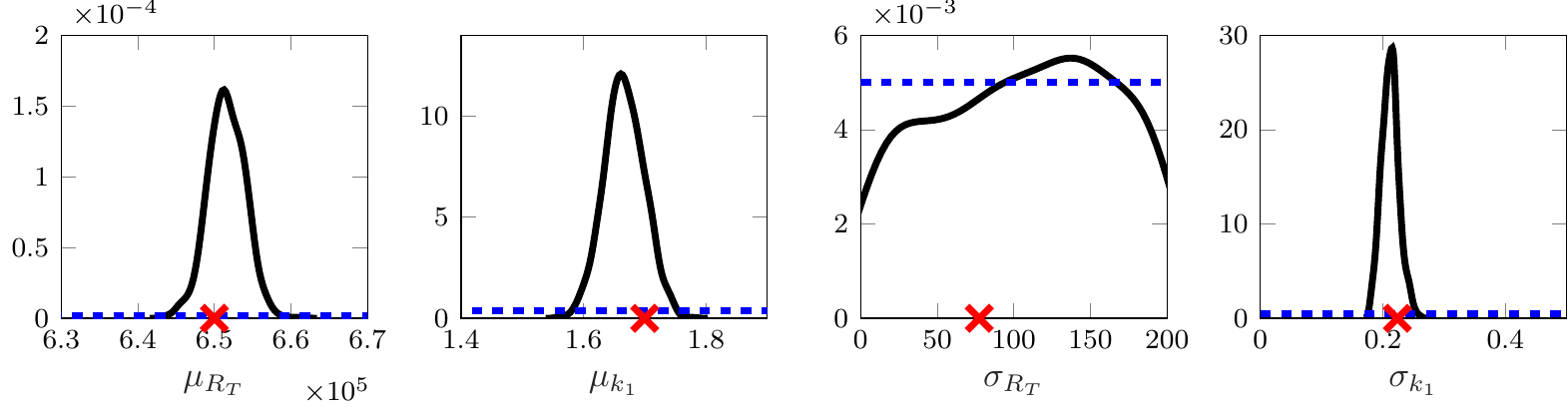}\label{figsub:growth_posterior_theta}}
		\subfigure[]{\includegraphics{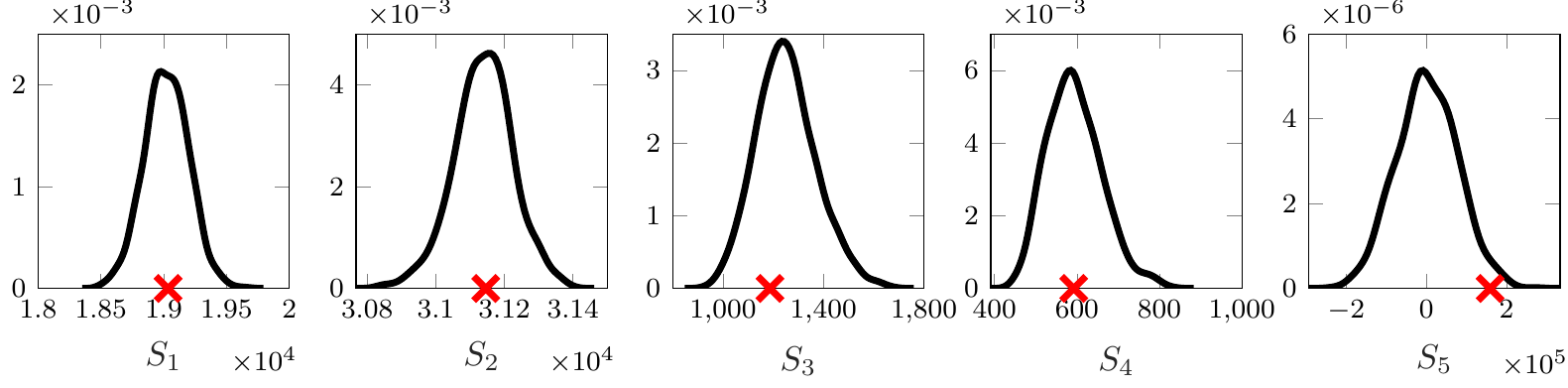}\label{figsub:growth_posterior_output}}
		\subfigure[]{\includegraphics{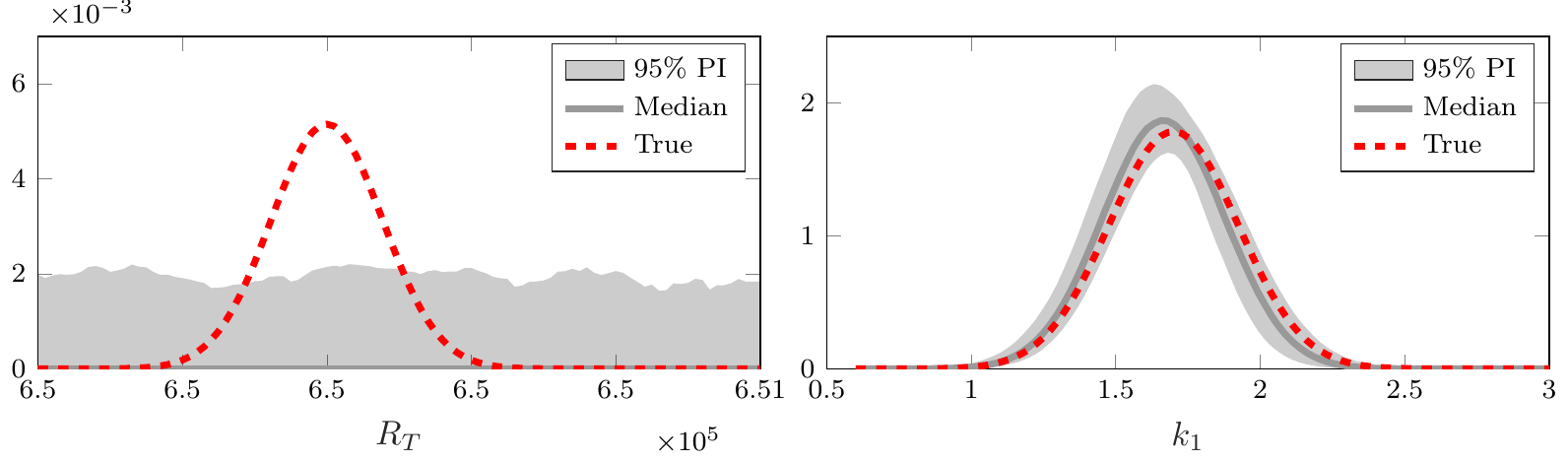}\label{figsub:growth_posterior_input}}
		\subfigure[]{\includegraphics{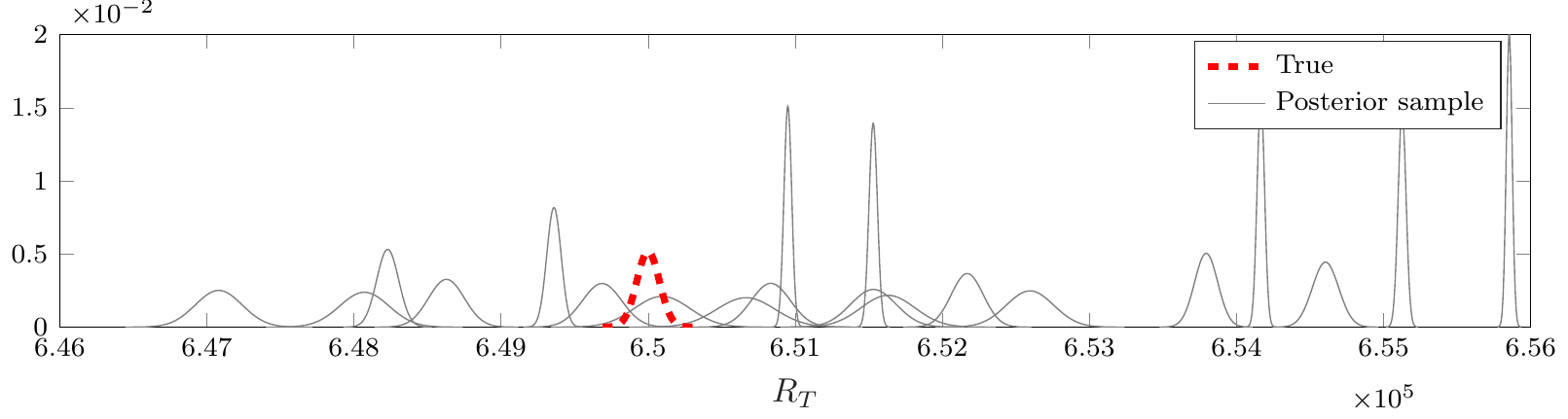}\label{figsub:growth_posterior_sims_input}}
		
		\caption{Results for the growth example with two unknown model parameters. (a) Univariate posterior distributions of $\theta$ with true values shown. (b) Posterior predictive distribution of the summaries of $h(y)$ with the the summaries of the population data overlaid.  (c) Posterior distribution (median and 95\% PI) of $f(x)$ with the true $f(x)$ overlaid for $R_T$ (left) and $k_1$ (right). (d) A selection of posterior samples of $f(x)$ for $R_T$ (grey solid), demonstrating the variety of permissible distributions for this parameter. The true $f(x)$ for $R_T$ is overlaid (red dashed).}
		\label{fig:growth_example}
	\end{figure}

	\begin{figure}[!tp]
		\centering
		
		\subfigure[]{\includegraphics{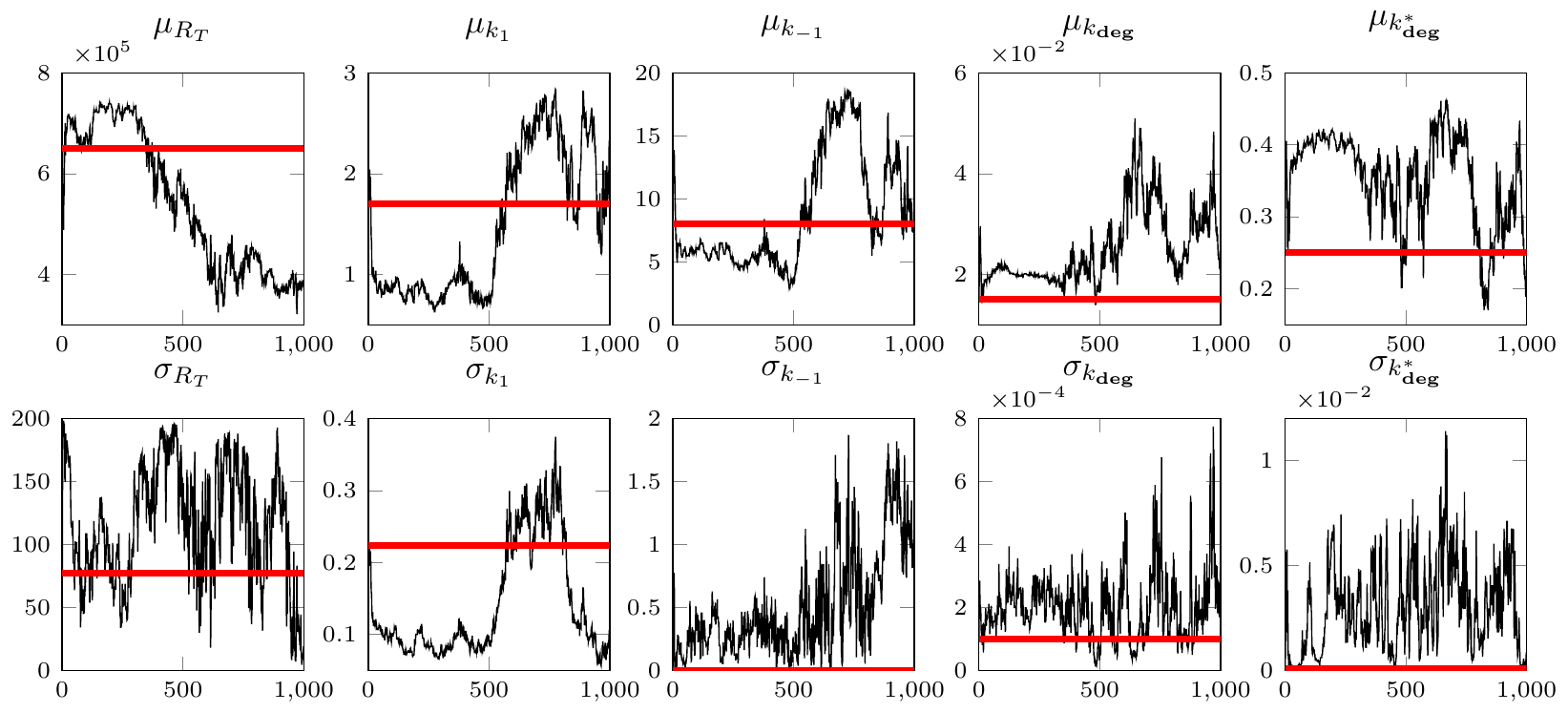}\label{figsub:growth_theta_trace_5param}}
		\subfigure[]{\includegraphics{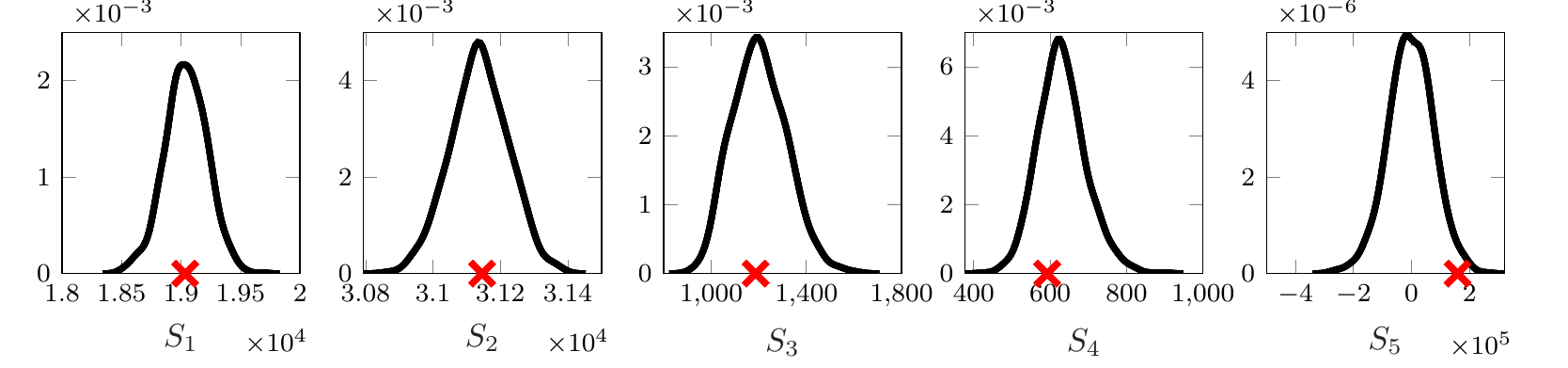}\label{figsub:growth_posterior_output_5param}}
		\subfigure[]{\includegraphics{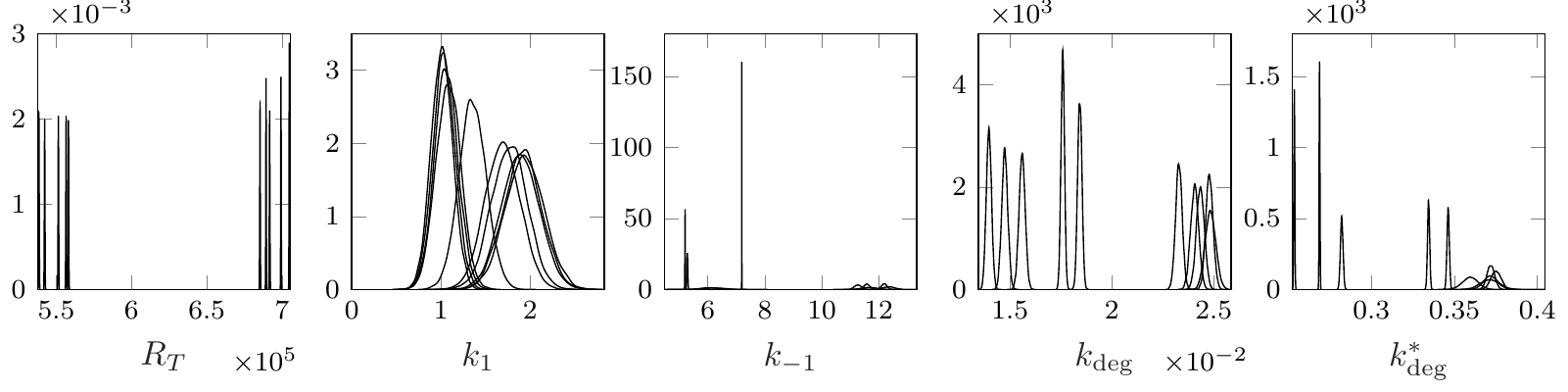}\label{figsub:growth_input_sample_5param}}
		
		\caption{Results for the growth example with five unknown model parameters.  (a) MCMC trace plots of $\theta$ with true value shown as a horizontal red solid line. (b) Posterior predictive distribution of the summaries the $h(y)$ with the the summaries of the population data overlaid. (c) Sample of 10 population distributions of model parameters obtained from MCMC.}
		\label{fig:growth_example_5param}
		
	\end{figure}
	
	Here we consider population calibration for the deterministic growth factor model considered in \citet{Dixit2020} and \citet{Lambert2021}.  Consider the following coupled ordinary differential equations:
	\begin{align*}
	\frac{dR}{dt} &= R_T k_{\mathrm{deg}} - k_1LR(t) + k_{-1}P(t) - k_{\mathrm{deg}} R(t), \\
	\frac{dP}{dt} &=  k_1LR(t) - k_{-1}P(t) - k_{\mathrm{deg}}^* R(t),
	\end{align*}
	where $R(t)$ and $P(t)$ are the amount of ligand-free and ligand-bound receptors on the cell surface, respectively. The ligand is exogenously supplied, and its amount is treated as a fixed quantity, $L$.  We denote the model parameter as $x = (R_T, k_1, k_{-1}, k_{\mathrm{deg}}, k_{\mathrm{deg}}^*)$.  Denote $P(t;x,L)$ as the value of $P(t)$ for a given $x$ and $L$ as found by forwards solution of the differential equation system.  We consider a similar set-up to \citet{Lambert2021}, fixing $(k_{-1}, k_{\mathrm{deg}}, k_{\mathrm{deg}}^*) = (8, 0.015, 0.25)$ and drawing $R_T \sim \mathcal{N}(6.5 \times 10^{5}, 0.6 \times 10^4)$, $k_1 \sim \mathcal{N}(1.7, 0.05)$.  Bivariate population data is produced by observing the level of ligand-bound receptors for two different levels of ligand supply, across many different cells exhibiting the specified heterogeneity in parameters $R_T$ and $k_1$. That is, we make independent draws of $P(10;x,2), P(10;x',10)$ where $x'$ denotes an independent replicate of $x$.  As noted in \citet{Lambert2021}, the underlying population distribution $h(y)$ can be approximated by a bivariate normal distribution with zero correlation.
	
	We assume that $n=100$ samples are available from $h(y)$.  We summarise the data with the two univariate sample means and variances, and the covariance between the two components, creating a five-dimensional summary statistic.  For simplicity we denote these summaries as $S_1,\ldots,S_5$. We attempt to infer the underlying distributions of $R_T$ and $k_1$ that when pushed through the deterministic model can recapture the observed population data, or at least its features.  We assume that distributions of $R_T$ and $k_1$ are correctly specified as independent Gaussians and we attempt to infer the means and standard deviations, $\theta = (\mu_{R_T}, \mu_{k_1}, \sigma_{R_T}, \sigma_{k_1})$.  For the prior we use independent uniform distributions, $\mu_{R_T} \sim \mathcal{U}(2.5 \times 10^5, 8 \times 10^5)$, $\mu_{k_1} \sim \mathcal{U}(0.25, 3)$, $\sigma_{R_T} \sim \mathcal{U}(0, 200)$, $\sigma_{k_1} \sim \mathcal{U}(0, 2)$, which are specifications that are guided by \citet{Lambert2021} .  We use 20000 iterations of MCMC BSL to infer the posterior distribution of $\theta$. 
	
	The results are shown in Figure \ref{fig:growth_example}.  Figure \ref{fig:growth_example}(a)  shows the estimated univariate posterior distributions of $\theta$.  It can be seen that the components of $\theta$ are well identified, except for $\sigma_{R_T}$, which has a posterior distribution similar to its prior.  Figure \ref{fig:growth_example}(b)  shows, unsurprisingly, that the model is able to capture accurately the observed summary statistics of the population data.  Features of the posterior distribution for the distribution of $R_T$ and $k_1$ throughout the population are shown in Figure \ref{fig:growth_example}(c).  It is evident that the population distribution of $k_1$ is tightly constrained by the summaries of the population data.  This is not surprising given the data is highly informative about the distribution's mean and standard deviation (Figure \ref{fig:growth_example}(a))  The posterior median is close to the true distribution and the 95\% CI closely envelopes the latter.  In contrast, the population distribution of $R_T$ is not well identified from the population data.  Even though the mean of $R_T$ is somewhat well identified, there is a great deal of uncertainty in its standard deviation (\ref{fig:growth_example}(a)).  This implies that a wide variety of distributions on $R_T$ can recover the summary statistics of the population data.  Such distributions are shown in Figure \ref{fig:growth_example}(d).    
	
	Now we demonstrate how our method can be practically difficult to work with.  Using the same dataset as above, we consider the same problem in \citet{Lambert2021} where the marginal population distribution of all five model parameters is calibrated.  For our approach, $\theta$ now consists of 10 hyperparameters, $\theta = (\mu_{R_T}, \mu_{k_1}, \mu_{k_{-1}}, \mu_{k_{\mathrm{deg}}}, \mu_{k_{\mathrm{deg}}^*}, \sigma_{R_T}, \sigma_{k_1}, \sigma_{k_{-1}}, \sigma_{k_{\mathrm{deg}}}, \sigma_{k_{\mathrm{deg}}^*})$.   For the prior we use independent uniform distributions, $\mu_{R_T} \sim \mathcal{U}(2.5 \times 10^5, 8 \times 10^5)$, $\mu_{k_1} \sim \mathcal{U}(0.25, 3)$,  $\mu_{k_{-1}} \sim \mathcal{U}(2, 20)$,   $\mu_{k_{\mathrm{deg}}} \sim \mathcal{U}(0.005, 0.1)$,  $\mu_{k_{\mathrm{deg}}^*} \sim \mathcal{U}(0.1, 0.5)$, $\sigma_{R_T} \sim \mathcal{U}(0, 200)$, $\sigma_{k_1} \sim \mathcal{U}(0, 2)$, $\sigma_{k_{-1}} \sim \mathcal{U}(0, 2)$, $\sigma_{k_{\mathrm{deg}}} \sim \mathcal{U}(0, 2)$, $\sigma_{k_{\mathrm{deg}}^*} \sim \mathcal{U}(0, 2)$.  If the data are informative enough, we should find that for $k_{-1}, k_{\mathrm{deg}}, k_{\mathrm{deg}}^*$, the estimated $\mu$ is close to the true parameter value and $\sigma$ is estimated to be close to zero.

	After several pilot runs to help inform the proposal distribution, we use 50000 iterations of MCMC BSL, producing trace plots as shown in Figure \ref{fig:growth_example_5param}(a).  It is evident that the MCMC has failed to converge.  We find that the posterior distribution of $\theta$ is difficult to explore, with highly complex dependencies between hyperparameters \emph{a posteriori} (results not shown).  Despite this, Figure \ref{fig:growth_example_5param}(b) shows that the model is capable of recovering the observed summary statistics of the data, and is very similar to Figure \ref{fig:growth_example}(b) for the two parameter problem.  Trace plots for these summary statistics are also quite reasonable (results not shown).  These results demonstrate that there are a huge variety of populations, differing in the nature of their heterogeneity, that all produce the observed variability in the data for ligand-bound receptor levels. This variety is further evident in Figure \ref{fig:growth_example_5param}(c), where 10 randomly generated population distributions of the model parameters sampled from the MCMC chain are shown.  The failure to converge truly highlights the hidden difficulty of population calibration problems for even deterministic models, especially when taking a fully Bayesian approach as we have done here.  In such cases where the full population posterior is too difficult or expensive to appropriately explore, methods returning a single distribution that has been appropriately regularised do become more appealing.  However, the approach of \citet{Lambert2021} to this problem significantly misspecifies the values (and distributions) of several of the model parameters, as compared to the true values used to generate the population data.  An advantage of our approach is that it is able to quantify how informative a set of population data is.

	\subsection{Flow Cytometry}
	
	To demonstrate population calibration using real and potentially noisy experimental data, here we consider population variability in the \textit{internalisation} of material by cells, where data are collected using flow cytometry \citep{Browning2021}.  A distinguishing feature of this problem is the quantity of data available: flow cytometry provides single-cell information at rates exceeding several thousand cells per second \citep{ONeill.2013}, so our data comprise samples from several million cells \citep{Browning2021}. In the experiments, cells are incubated with fluorescent-labelled antibody that is internalised by transferrin receptors, the route usually reserved for the uptake of iron. The population calibration problem here is to infer distributional information relating to the variation in internalisation rates and cell size.  We previously apply our approach to this population calibration problem in \citet{Browning2021} with code and data available at \url{https://github.com/ap-browning/internalisation}.  The main focus of that paper is on the new biological insight that can be gained by considering population rather than averaged data.  Here we focus mainly on the mathematical and statistical aspects of the problem, and refer to \citet{Browning2021} for more biological detail and motivation.

	As the number of receptors and antibody molecules present on each cell is relatively large ($\sim 10^5$), the transient dynamics can be modelled using the system of ODEs
	\begin{equation}\label{eq:SysODEs}
	\left.\begin{aligned}
	\dv{T(t)}{t} &= -\beta T(t),\\
	\dv{S(t)}{t} &= \beta T(t) - \lambda S(t) + p \beta E(t),\\
	\dv{E(t)}{t} &= \lambda S(t) - p\beta E(t),\\
	\dv{F(t)}{t} &= p\beta E(t).
	\end{aligned}\right.
	\end{equation}
	Here, $T(t)$ represents the relative number of internalised transferrin-bound receptors; $S(t)$ that of antibody-bound receptors on the cell surface; $E(t)$ that of antibody-bound receptors inside the cell; $\lambda$ [\SI{}{\per\minute}] is the rate at which antibody-bound receptors are internalised; and, $\beta$ [\SI{}{\per\minute}] is the rate at which internalised receptors recycle to the cell surface. As the number of receptors present on each cell remains approximately constant, yet the fluorescent signal increases, we assume that a small proportion of internalised receptor-bound antibody, $p E(t)$, disassociates and accumulates in a pool $F(t)$ of free, internalised, antibody. All molecule counts are modelled as proportions of the total number of receptors $R = T(0) + S(0)$ present in each cell, and initially the system is assumed to be in an antibody-free equilibrium \citep{Browning2021}.
	
	Flow cytometry \citep{ONeill.2013} is used to collect noisy, single-cell, measurements relating to the amount of fluorescent material (i.e., antibody) present in each cell. To obtain information relating to both the total amount of antibody, $A(t) = S(t) + E(t) + F(t)$ and amount internalised on each cell, $I(t) = E(t) + F(t)$, antibody are labelled with two fluorescent probes, one of which can be selectively switched-off through the introduction of a quencher dye. Flow cytometry measurements typically comprise two sources of error: (1) circuit-derived noise, which can be modelled as Gaussian with variance proportional to the signal \citep{Galbusera.2020}; and (2) cellular autofluorescence, or background fluorescence, which we model by constructing an empirical distribution using samples without antibody.
	
	Measurements from each cell are, therefore, modelled by
	\begin{align}\label{eq:flowcytometrymodel_unquenched}
	\begin{bmatrix} M_1 \\ M_2 \end{bmatrix} &= 
	\left\{\begin{array}{ll}
	\begin{bmatrix}
	\alpha_1 A(t) R + \sqrt{\alpha_1 A(t) R} \sigma_1\varepsilon_1 + E_1\\
	\alpha_2 A(t) R + \sqrt{\alpha_2 A(t) R} \sigma_2\varepsilon_2 + E_2
	\end{bmatrix} & 
	\text{No quencher dye,}\\\vspace{0.05cm}\\
	\begin{bmatrix}
	\alpha_1 \tilde{I}(t) R + \sqrt{\alpha_1 \tilde{I}(t) R} \sigma_1\varepsilon_1 + E_1\\
	\alpha_2 A(t) R + \sqrt{\alpha_2 A(t) R} \sigma_2\varepsilon_2 + E_2
	\end{bmatrix} & 
	\text{Quencher dye.}\\
	\end{array}\right.
	\end{align}
	Here, $\tilde{I}(t) = I(t) + (1 - \eta) S(t)$, where $\eta$ is the \textit{quenching efficiency}; that is, the probability that the fluorescence of a surface-bound antibody molecule is switched off by the introduction of the quenched dye. $\eta \approx 0.94$ is pre-estimated from the data; $\alpha_1$ and $\alpha_2$ are proportionality constants, relating the amount of fluorescent molecules to the magnitude of the flow cytometry signal; $\varepsilon_1,\varepsilon_2 \sim \mathcal{N}(0,1)$; and, $(E_1,E_2)$ are jointly distributed random variables that represent cellular autofluorescence.
	
	We capture cell-to-cell variability by modelling the number of receptors present in each cell, $R$, the internalisation rate, $\lambda$, and the recycling rate, $\beta$ as jointly distributed random variables $\bm\xi_i = (R_i,\lambda_i,\beta_i)$ where here the subscript $i$ refers to the $i$th cell. Without loss of generality, we set $\mathbb{E}(R) = 1$ (i.e., receptor counts are relative to the population average). We assume that cell properties have a unimodal distribution and assume a parametric form of $\bm\xi_i$ that allows us to infer the first three moments. The marginals are given by
	\begin{equation}
	\begin{aligned}
	R_i &\sim \mathrm{ShiftedLogNormal}(\mu_R,\sigma_R),\\
	\lambda_i &\sim \mathrm{ShiftedGamma}(\mu_\lambda,\sigma_\lambda,\omega_\lambda),\\
	\beta_i &\sim \mathrm{ShiftedGamma}(\mu_\beta,\sigma_\beta,\omega_\beta),\\
	\end{aligned}
	\end{equation}	
	where $\lambda_i$ and $\beta_i$ are shifted Gamma variables parameterised in terms of their respective means, standard deviations and skewnesses (negative skewness is allowed). Consistent with existing statistical modelling of flow cytometry data \citep{Furusawa.2005}, $R_i$ is assumed to be log-normally distributed. To maintain positivity, we truncate all distributions so that $R_i,\lambda_i,\beta_i > 0$ (variables are parameterised in terms of the untruncated distributions). To allow for correlations between cell properties, we model the joint distribution of $\bm\xi_i$ with a Gaussian copula with covariance matrix
	\begin{equation}
	\mathbf{P} = \begin{pmatrix}%
	1 & \rho_{R\lambda} & \rho_{R\beta}\\
	\rho_{R\lambda} & 1 & \rho_{\lambda\beta}\\
	\rho_{R\beta} & \rho_{\lambda\beta} & 1
	\end{pmatrix}.
	\end{equation}
	Here, $\rho_{R\beta} = \rho_{R\lambda} \rho_{\lambda\beta} + \tilde{\rho}_{R\beta}\sqrt{(1 - \rho_{R\lambda}^2)(1 - \rho_{\lambda\beta}^2)}$ for $ \tilde{\rho}_{R\beta} \in (0,1)$ such that $\mathbf{P}$ remain positive definite.
	
\begin{figure}
		\centering
		
		\subfigure[]{\includegraphics{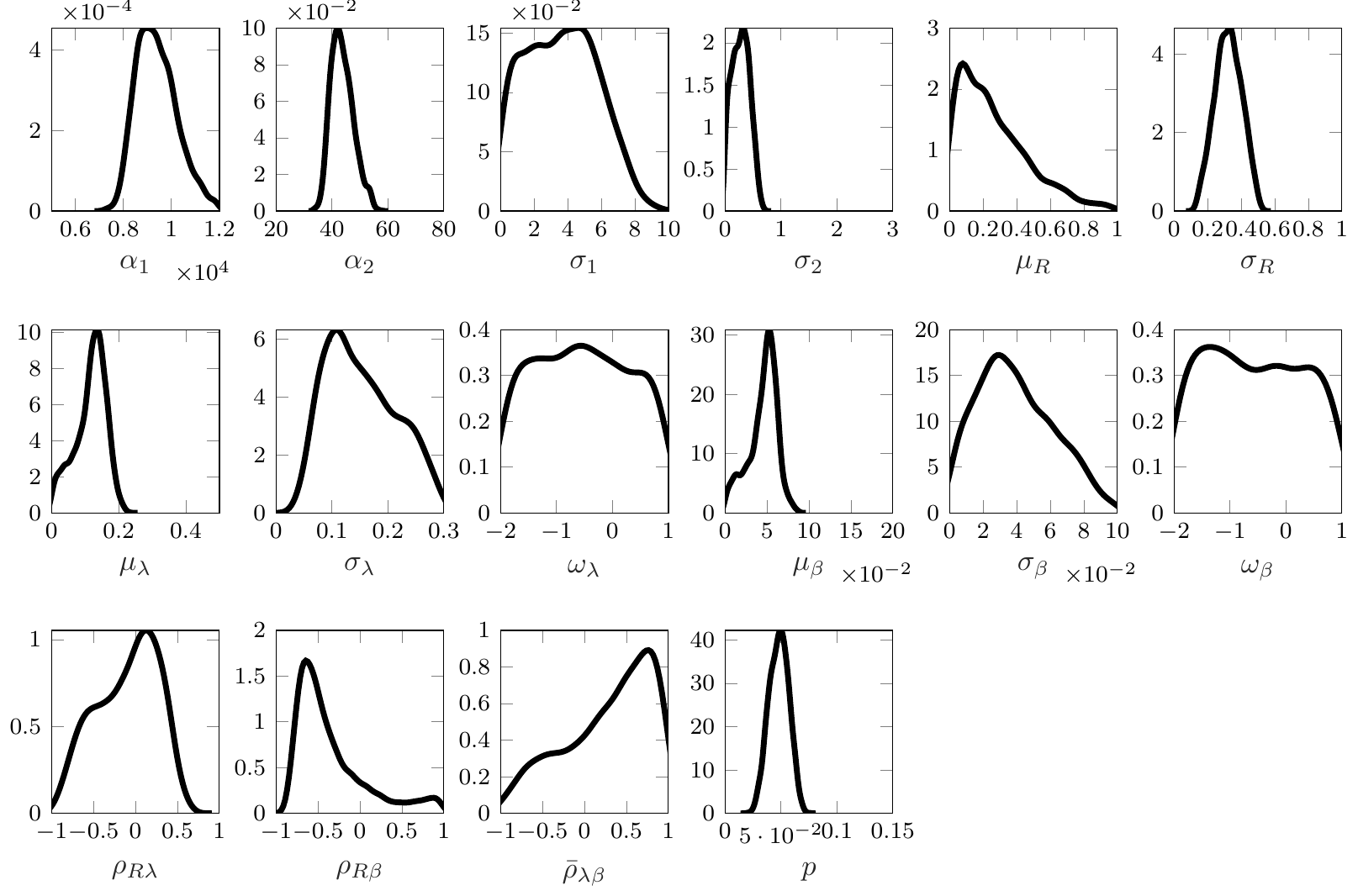}\label{figsub:internalisation_posterior}}
		\subfigure[]{\includegraphics{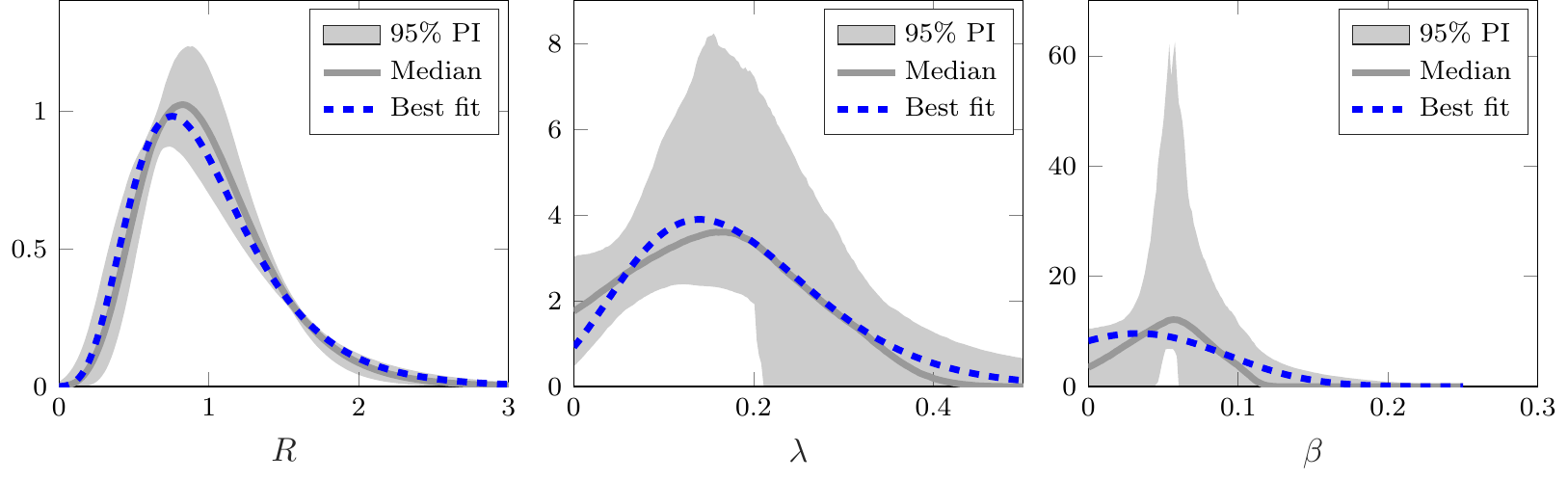}\label{figsub:internalisation_predictive}}
		
		\caption{Results for the internalisation model. (a) Univariate posterior distributions for $\theta$ and $\phi$. Uniform priors are used, with support corresponding to respective axis limits. (b) Posterior distribution of $f(x)$, $x = (R, \lambda,\beta)$, shown as median with 95\% PI (grey). A best fit distribution was obtained as the posterior sample with the lowest average discrepancy from 100 simulations (blue dashed).}
		\label{fig:internalisation_posterior}
	\end{figure}

	Compared to the previous case studies, the internalisation model is heavily parameterised; $\theta = (\mu_R,\sigma_R,\mu_\lambda,\sigma_\lambda,\omega_\lambda,\mu_\beta,\sigma_\beta,\omega_\beta,\rho_{R\lambda},\rho_{R\beta},\tilde{\rho}_{R\beta})$ and $\phi = (\alpha_1, \alpha_2, \sigma_1,\sigma_2,p)$. We assume uniform priors (bounds correspond to the axis limits in Figure \ref{fig:internalisation_posterior}(a)) and infer the posterior distribution of $(\theta,\phi)$ using a hybrid SMC/MCMC ABC approach based on a discrepancy metric that matches a weighted sum of the correlations between measurements $M_1$ and $M_2$, and the marginal distributions of each using the Anderson-Darling distance
	\begin{equation}
	\mathrm{addist}^2(\mathsf{y},\mathsf{z}) = N \int_{-\infty}^\infty \dfrac{\left(F_\mathsf{z}(w) - F_\mathsf{y}(w)\right)^2}{F_\mathsf{y}(w)\big(1 - F_\mathsf{y}(w)\big)} \mathrm{d} F_\mathsf{y}(w),
	\end{equation}
	where $F_\mathsf{y}(w)$ is the distribution function for the observed data and $F_\mathsf{z}(w)$ is the empirical distribution function of $n$ synthetically generated observations. For full details, see \citet{Browning2021}. While the experimental data comprise $\sim 10^5$ cells per time point, quenched or not quenched, we simulate synthetic data sets based on $n = 10^3$ (we treat this as a tuning parameter to balance computational burden and the ABC acceptance rate).
	
	A major difficultly in the application of ABC is determining an appropriate acceptance tolerance. To address this, we first use an adaptive SMC ABC \citep{Vo2015} algorithm to identify the region of the parameter space with non-negligible posterior density, and establish an achievable acceptance tolerance. We then choose an ABC acceptance tolerance based on the particle with the smallest discrepancy identified with SMC ABC having an acceptance rate of 50\%. Next, we sample from the ABC posterior using 4 chains of $10^7$ iterations of MCMC ABC \citep{Marjoram2003}, thinning to every 10,000 samples. Posterior distributions are shown in Figure \ref{fig:internalisation_posterior}(a) and posterior predictive inputs are shown in Figure \ref{fig:internalisation_posterior}(b).
	
\begin{figure}
		\centering
		\includegraphics{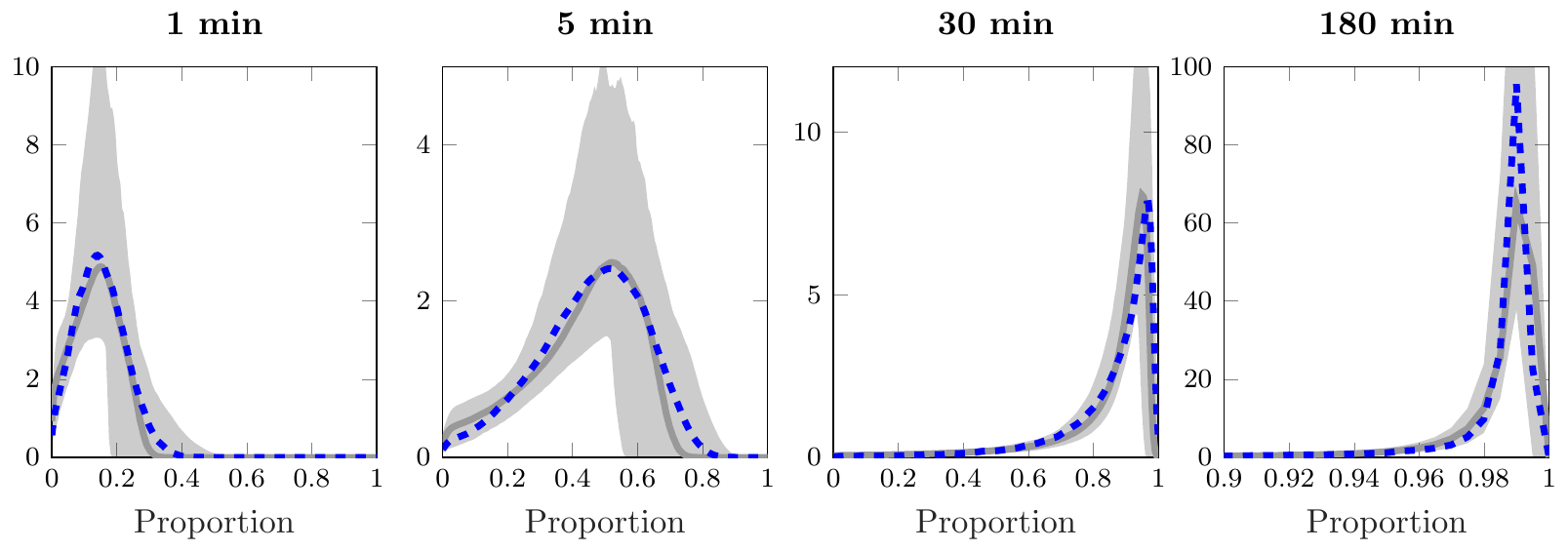}
		
		\caption{Posterior predictive distribution of the proportion of material internalised, $(E(t) + F(t))/(S(t) + E(t) + F(t))$. We show the median with 95\% PI (grey) and predictive distribution at the model best fit (blue dashed).}
		\label{fig:internalisation_predictive_proportion}
	\end{figure}

	Despite the relatively large sample size, many parameters are practically non-identifiable. In particular, results for $\omega_\lambda$ and $\omega_\beta$ indicate that model outputs are relatively insensitive to the shape of the internalisation and recycling rate distributions. However, we are able to identify that regions of the posterior where the internalisation rate is deterministic (i.e., $\sigma_\lambda = 0$) have negligible support. Further, estimates for the proportion of receptors that recycle, $p$, and estimates relating to the marginal distribution of $R$ (Figure \ref{fig:internalisation_posterior}(b)) are relatively constrained. Despite the lack of practical identifiability in $\theta$, in Figure \ref{fig:internalisation_predictive_proportion} we show that model predictions are relatively tightly constrained by producing posterior predictive distributions of the proportion of material internalised over the course of the experiment, a measurement that cannot be observed experimentally.

	\section{Discussion} \label{sec:Discussion}
	
	In this paper we have presented a solution to the population calibration problem using likelihood-free inference.  The advantages of the approach are that it can be applied, at least in principle, to any population calibration problem, and it produces uncertainty quantification on the estimated distribution of the model parameters.   As illustrated in the paper, the main limitation is computational.  Firstly, the approach requires a huge number of model solves or simulations.  Secondly, the method requires sampling over the space of hyperparameters that may have a much larger dimension than the model parameters. 
	
	The most similar approach to ours \citep{Hasenauer2011} uses in some sense a non-parametric approach where a mixture distribution with a relatively large number of components is used for $f(x)$ where the hyperparameters of each component is fixed and only the mixture weights are inferred.  This specification has the major advantage that model solves based on parameter values generated from the distribution of each mixture component can be generated and stored prior to running MCMC, and be recycled for different proposed mixture weights during MCMC.  Their approach is designed for a deterministic model with known noise distribution, but could be adapted to a more general context.  However, as the dimension of $x$ increases it is less clear how to set the hyperparameters of each mixture component.  There may be other ways to recycle or exploit model solutions/simulations that have been generated offline or throughout the MCMC.  For example, a surrogate model~\citep{Sacks1989} could be used to address the large number of forward simulations required by a fully Bayesian approach to population calibration if each is computationally costly. Even for problems too complex for construction of a sufficiently faithful surrogate, evaluations of the surrogate can still be incorporated into a sampling routine to improve proposals, and hence obtain speed-up without incurring any approximation error~\citep{Bon2021}.
	
	Another way to speed up the computations would be to replace MCMC with a variational approximation, which have been considered before in the likelihood-free literature (e.g. \citet{Tran2017,Ong2018b}).  This approach would only provide an approximation to the posterior distribution of the hyperparameters, but still may provide valuable insight into the uncertainty quantification.  We leave the exploration of more computationally efficient approaches for future research.   
	
	One potential issue that does not appear to have been addressed in the population calibration literature is model misspecification.  Model misspecification in the context of population calibration could imply that there is no distribution on model parameters that can reproduce the population data.  Fortunately, there has been some research in the likelihood-free literature for more standard Bayesian calibration in the presence of model misspecification, and the consequences that can arise \citep{frazier2020model,Frazier2021}.  \citet{frazier2019robust} extend BSL so that it can handle the situation where the model cannot reproduce all the summary statistics of the data, and can identify the offending summaries which can inform further model development.  See \citet{frazier2020robust} for an analogous approach to ABC.  Such methods could be adapted to the population calibration context, but we leave that for further research.    
	
	Population calibration problems make up a growing proportion of the biological literature, as the need for mechanistic approaches to accommodate variability becomes better understood (see for example~\citet{brown2015trauma,fuertinger2018virtual,jenner2021covid,jenner2021silico,cassidy2019determinants,barish2017evaluating,Britton2013}).  In particular, virtual clinical trials~\citep{alfonso2020translational} are becoming a popular way to examine an experimental treatment's robustness before moving forward to clinical trial~\citep{jenner2021covid,cassidy2019determinants,barish2017evaluating}. Appropriate statistical treatment of these problems has been demonstrated to produce more predictive virtual populations, and to better distinguish parameter differences underlying separate cohorts in a dataset~\citep{Lawson2018}. With the methodology presented here, these benefits now become available for models without an available likelihood, and we also gain a sense for the level of certainty in the population's predicted variability. Our hope is that these methods be more widely applied to variable population data, such as flow cytometry measurements, genomic data, and survival curves, and also further developed to reduce the computational burden.  Correctly capturing and characterising this variability, whether on the cellular level or the human level, is critical for these {\it in silico} studies to provide more reliable predictions and insights.

\section*{Acknowledgements}

CD was supported by the Australian Research Council. ALJ was supported by a Queensland University of Technology Early Career Researcher Scheme. APB was supported by an ARC Centre of Excellence for Mathematical and Statistical Frontiers Research SPRINT scheme and an Australian Mathematical Society Lift-Off Fellowship.
	
	\bibliographystyle{apalike} 
	\bibliography{refs}

\end{document}